# A Coordinated Approach to Channel Estimation in Large-scale Multiple-antenna Systems

Haifan Yin, David Gesbert *Fellow, IEEE*, Miltiades Filippou, and Yingzhuang Liu

*Abstract*—This paper addresses the problem of channel estimation in multi-cell interference-limited cellular networks. We consider systems employing multiple antennas and are interested in both the finite and large-scale antenna number regimes (so-called "massive MIMO"). Such systems deal with the multi-cell interference by way of per-cell beamforming applied at each base station. Channel estimation in such networks, which is known to be hampered by the pilot contamination effect, constitute a major bottleneck for overall performance. We present a novel approach which tackles this problem by enabling a low-rate coordination between cells during the channel estimation phase itself. The coordination makes use of the additional second-order statistical information about the user channels, which are shown to offer a powerful way of discriminating across interfering users with even strongly correlated pilot sequences. Importantly, we demonstrate analytically that in the large-number-of-antennas regime, the pilot contamination effect is made to vanish completely under certain conditions on the channel covariance. Gains over the conventional channel estimation framework are confirmed by our simulations for even small antenna array sizes.

## I. INTRODUCTION

Full reuse of the frequency across neighboring cells leads to severe interference, which in turn limits the quality of service offered to cellular users, especially those located at the cell edge. As service providers seek some solutions to restore performance in low-SINR cell locations, several approaches aimed at mitigating inter-cell interference have emerged in the last few years. Among these, the solutions which consist of exploiting the additional degrees of freedom made available by the use of multiple antennas seem the most promising, particularly so at the base station side where such arrays are more easily affordable.

In an effort to solve this problem while limiting the requirements for user data sharing over the backhaul network, coordinated beamforming approaches have been proposed in which 1) multiple-antenna processing is exploited at each base station, and 2) the optimization of the beamforming vectors at all cooperating base stations is performed jointly. Coordinated beamforming does not require the exchange of user message information (e.g., in network MIMO). Yet it still demands the exchange of channel state information (CSI) across the transmitters on a fast time scale, low-latency basis, making almost as challenging to implement in practice as the above mentioned network MIMO schemes.

Fortunately a path towards solving some of the essential practical problems related to beamforming-based interference avoidance was suggested in [1]. In this work, it was pointed out that the need for exchanging Channel State Information at Transmitter (CSIT) between base stations could be alleviated by simply increasing the number of antennas, $M$, at each transmitter (so-called massive MIMO). This result is rooted in the law of large numbers, which predicts that, as the number of antennas increases, the vector channel for a desired terminal will tend to be more orthogonal to the vector channel of a randomly selected interfering user. This makes it possible to reject interference at the base station side by simply aligning the beamforming vector with the desired channel ("Maximum Ratio Combining" or spatial matched filter). Hence in theory, a simple fully distributed per-cell beamforming scheme can offer performance scaling (with $M$) similar to a more complex centralized optimization.

Unfortunately, the above conclusion only holds without pilot-contaminated CSI estimates. In reality, channel information is acquired on the basis of finite-length pilot sequences, and crucially, in the presence of inter-cell interference. Therefore, the pilot sequences from neighboring cells would contaminate each other. It was pointed out in [1] that pilot contamination constitutes a bottleneck for performance. In particular, it has been shown that pilot contamination effects [2], [3], [4] (i.e., the reuse of non-orthogonal pilot sequences across interfering cells) cause the interference rejection performance to quickly saturate with the number of antennas, thereby undermining the value of MIMO systems in cellular networks.

In this paper, we address the problem of channel estimation in the presence of multi-cell interference generated from pilot contamination. We propose an estimation method which provides a substantial improvement in performance. It relies on two key ideas. The first is the exploitation of dormant side-information lying in the second-order statistics of the user channels, both for desired and the interfering users. In particular, we demonstrate a powerful result indicating that the exploitation of covariance information under certain subspace conditions on the covariance matrices can lead to a complete removal of pilot contamination effects in the large $M$ limit. We then turn to a practical algorithm design where this concept

Manuscript received February 1, 2012; revised June 15, 2012. This work was supported in part by National Natural Science Foundation of China under grant No. 60972015 and 61231007, by the European research project SAPHYRE under the FP7 ICT Objective 1.1 - The Network of the Future, and by the French national ANR-VERSO funded project LICORNE.

H. Yin was with Huazhong University of Science and Technology, Wuhan, 430074 China. He is now with EURECOM, 06410 Biot, France (e-mail: yin@eurecom.fr).

D. Gesbert, M. Filippou are with EURECOM, 06410 Biot, France (e-mail: gesbert@eurecom.fr, filippou@eurecom.fr).

Y. Liu is with Huazhong University of Science and Technology, Wuhan, 430074 China (e-mail: liuyz@mail.hust.edu.cn).

4is exploited. The key idea behind the new algorithm is the use of a covariance-aware pilot assignment strategy within the channel estimation phase itself. While diversity-based scheduling methods have been popularized for maximizing various throughput-fairness performance criteria [5], [6], [7], [8], the potential benefit of user-to-pilot assignment in the context of interference-prone channel estimation has received very little attention so far.

More specifically, our contributions are the following: We first develop a Bayesian channel estimation method making explicit use of covariance information in the inter-cell interference scenario with pilot contamination. We show that the channel estimation performance is a function of the degree to which dominant signal subspaces pertaining to the desired and interference channel covariance overlap with each other. Therefore we exploit the fact that the desired user signals and interfering user signals are received at the base station with (at least approximately) finite-rank covariance matrices. This is typically the case in realistic scenarios due to the limited angle spread followed by incoming paths originating from street-level users [9]. Finally, we propose a pilot sequence assignment strategy based on assigning carefully selected groups of users to identical pilot sequences. The gains are shown to depend on system parameters such as the typical angle spread measured at the base station and the number of base station antennas. Performance close to the interference-free channel estimation scenario is obtained for moderate numbers of antennas and users.

The notations adopted in the paper are as follows. We use boldface to denote matrices and vectors. Specifically, $\mathbf{I}_M$ denotes the $M \times M$ identity matrix. Let $(\mathbf{X})^T$, $(\mathbf{X})^*$, and $(\mathbf{X})^H$ denote the transpose, conjugate, and conjugate transpose of a matrix $\mathbf{X}$ respectively. $\mathbb{E}\{\cdot\}$ denotes the expectation, $\|\cdot\|_F$ denotes the Frobenius norm, and $\mathrm{diag}\{\mathbf{a_1}, ..., \mathbf{a_N}\}$ denotes a diagonal matrix or a block diagonal matrix with $\mathbf{a_1}, ..., \mathbf{a_N}$ at the main diagonal. The Kronecker product of two matrices $\mathbf{X}$ and $\mathbf{Y}$ is denoted by $\mathbf{X} \otimes \mathbf{Y}$. $\triangleq$ is used for definitions.

## II. SIGNAL AND CHANNEL MODELS

We consider a network of $L$ time-synchronized[1] cells, with full spectrum reuse. Estimation of (block-fading) channels in the uplink is considered,[2] and all the base stations are equipped with $M$ antennas. To simplify the notations, we assume the 1st cell is the target cell, unless otherwise notified. We assume the pilots, of length $\tau$, used by single-antenna users in the same cell are mutually orthogonal. As a result, intra-cell interference is negligible in the channel estimation phase. However, non-orthogonal (possibly identical) pilots are reused from cell to cell, resulting in pilot contamination from $L-1$ interfering cells. For ease of exposition, we consider the case where a single user per cell transmits its pilot sequence to its serving base. The pilot sequence used in the $l$-th cell is denoted by:

$$\mathbf{s}_l = \begin{bmatrix} s_{l1} & s_{l2} & \cdots & s_{l\tau} \end{bmatrix}^T. \quad (1)$$

The power of pilot sequences are assumed equal such that $|s_{l1}|^2 + \cdots + |s_{l\tau}|^2 = \tau, l = 1, 2, \ldots, L$.

The channel vector between the $l$-th cell user and the target base station is $\mathbf{h}_l$. Thus, $\mathbf{h}_1$ is the desired channel while $\mathbf{h}_l$, $l > 1$ are interference channels. All channel vectors are assumed to be $M \times 1$ complex Gaussian, undergoing correlation due to the finite multipath angle spread at the base station side [10]:

$$\mathbf{h}_l = \mathbf{R}_l^{1/2} \mathbf{h}_{\mathrm{W}l}, l = 1, 2, \ldots, L, \quad (2)$$

where $\mathbf{h}_{\mathrm{W}l} \sim \mathcal{CN}(\mathbf{0}, \mathbf{I}_M)$ is the spatially white $M \times 1$ SIMO channel, and $\mathcal{CN}(\mathbf{0}, \mathbf{I}_M)$ denotes zero-mean complex Gaussian distribution with covariance matrix $\mathbf{I}_M$. In this paper, we make the assumption that covariance matrix $\mathbf{R}_l \triangleq \mathbb{E}\{\mathbf{h}_l \mathbf{h}_l^H\}$ can be obtained separately from the desired and interference channels (see Section VI for how this could be done in practice).

During the pilot phase, the $M \times \tau$ signal received at the target base station is

$$\mathbf{Y} = \sum_{l=1}^{L} \mathbf{h}_l \mathbf{s}_l^T + \mathbf{N}, \quad (3)$$

where $\mathbf{N} \in \mathbb{C}^{M \times \tau}$ is the spatially and temporally white additive Gaussian noise (AWGN) with zero-mean and element-wise variance $\sigma_n^2$.

## III. COVARIANCE-BASED CHANNEL ESTIMATION

### A. Pilot Contamination

Conventional channel estimation relies on correlating the received signal with the known pilot sequence (referred here as Least Squares estimate for example). Hence, using the model in (3), a Least Square (LS) estimator for the desired channel $\mathbf{h}_1$ is

$$\widehat{\mathbf{h}}_1^{\mathrm{LS}} = \mathbf{Y} \mathbf{s_1}^* (\mathbf{s_1}^T \mathbf{s_1}^*)^{-1}. \quad (4)$$

The conventional estimator suffers from a lack of orthogonality between the desired and interfering pilots, an effect known as pilot contamination [2], [11], [12]. In particular, when the *same* pilot sequence is reused in all $L$ cells, i.e., $\mathbf{s}_1 = \cdots = \mathbf{s}_L = \mathbf{s}$, the estimator can be written as

$$\widehat{\mathbf{h}}_1^{\mathrm{LS}} = \mathbf{h}_1 + \sum_{l \neq 1}^{L} \mathbf{h}_l + \mathbf{N} \mathbf{s}^*/\tau . \quad (5)$$

As it appears in (5), the interfering channels leak directly into the desired channel estimate. The estimation performance is then limited by the signal to interfering ratio at the base station, which in turns limits the ability to design effective interference-avoiding beamforming solution.

---

[1]Note that assuming synchronization between uplink pilots provides a worst case scenario from a pilot contamination point of view, since any lack of synchronization will tend to statistically decorrelate the pilots.

[2]Similar ideas would be applicable for downlink channel estimation, provided the UE is equipped with multiple antennas as well, in which case the estimation would help resolve interferences originating from neighboring base stations.

*B. Bayesian Estimation*

We hereby propose an improved channel estimator with the aim of reducing the pilot contamination effect, and taking advantage of the multiple antenna dimensions. We suggest to do so by exploiting side information lying in the second order statistics of the channel vectors. The role of covariance matrices is to capture structure information related to the distribution (mainly mean and spread) of the multipath angles of arrival at the base station. Due to the typically elevated position of the base station, rays impinge on the antennas with a finite angle-of-arrival (AOA) spread and a user location-dependent mean angle. Note that covariance-aided channel estimation itself is not a novel idea, e.g., in [13]. In [14], the authors focus on optimal design of pilot sequences and they exploit the covariance matrices of desired channels and colored interference. The optimal training sequences were developed with adaptation to the statistics of disturbance. In our paper, however, the pilot design is shown not having an impact on interference reduction, since fully aligned pilots are transmitted. Instead, we focus on i) studying the limiting behavior of covariance-based estimates in the presence of interference and large-scale antenna arrays, and ii) how to *shape* covariance information for the full benefit of channel estimation quality.

Two Bayesian channel estimators can be formed. In the first, all channels are estimated at the target base station (including interfering ones). In the second, only $\mathbf{h}_1$ is estimated. By vectorizing the received signal and noise, our model (3) can be represented as

$$\mathbf{y} = \tilde{\mathbf{S}}\mathbf{h} + \mathbf{n}, \quad (6)$$

where $\mathbf{y} = \text{vec}(\mathbf{Y})$, $\mathbf{n} = \text{vec}(\mathbf{N})$, and $\mathbf{h} \in \mathbb{C}^{LM \times 1}$ is obtained by stacking all $L$ channels into a vector. The pilot matrix $\tilde{\mathbf{S}}$ is defined as

$$\tilde{\mathbf{S}} \triangleq \begin{bmatrix} \mathbf{s}_1 \otimes \mathbf{I}_M & \cdots & \mathbf{s}_L \otimes \mathbf{I}_M \end{bmatrix}. \quad (7)$$

Applying Bayes' rule, the conditional distribution of the channels $\mathbf{h}$ given the received training signal $\mathbf{y}$ is

$$p(\mathbf{h}|\mathbf{y}) = \frac{p(\mathbf{h})p(\mathbf{y}|\mathbf{h})}{p(\mathbf{y})} = p(\mathbf{h})p(\mathbf{y}|\mathbf{h}). \quad (8)$$

We use the multivariate Gaussian probability density function (PDF) of the random vector $\mathbf{h}$ and assume its rows $\mathbf{h}_1, \cdots, \mathbf{h}_L$ are mutually independent, giving the joint PDF:

$$p(\mathbf{h}) = \frac{\exp\left(-\sum_{l=1}^{L} \mathbf{h}_l^H \mathbf{R}_l^{-1} \mathbf{h}_l\right)}{\pi^{LM}(\det \mathbf{R}_1 \cdots \det \mathbf{R}_L)^M}. \quad (9)$$

Note that we derive this Bayesian estimator under the standard condition of covariance matrix invertibility, although we show later this hypothesis is actually challenged by reality in the large-number-of-antennas regime. Fortunately, our final expressions for channel estimators completely skip the covariance inversion.

Using (6), we may obtain:

$$p(\mathbf{y}|\mathbf{h}) = \frac{\exp\left(-(\mathbf{y} - \tilde{\mathbf{S}}\mathbf{h})^H (\mathbf{y} - \tilde{\mathbf{S}}\mathbf{h})/\sigma_n^2\right)}{(\pi \sigma_n^2)^{M\tau}}. \quad (10)$$

Combining the equations (9) and (10), the expression of (8) can be rewritten as

$$p(\mathbf{h}|\mathbf{y}) = \frac{\exp(-l(\mathbf{h}))}{AB}, \quad (11)$$

where $A \triangleq (\pi \sigma_n^2)^{M\tau}$, $B \triangleq \pi^{LM}(\det \mathbf{R}_1 \cdots \det \mathbf{R}_L)^M = \pi^{LM}(\det \mathbf{R})^M$, and

$$l(\mathbf{h}) \triangleq \mathbf{h}^H \bar{\mathbf{R}} \mathbf{h} + (\mathbf{y} - \tilde{\mathbf{S}}\mathbf{h})^H (\mathbf{y} - \tilde{\mathbf{S}}\mathbf{h})/\sigma_n^2, \quad (12)$$

in which $\mathbf{R} \triangleq \text{diag}(\mathbf{R}_1, \cdots, \mathbf{R}_L)$, $\bar{\mathbf{R}} \triangleq \mathbf{R}^{-1}$.

Using the maximum a posteriori (MAP) decision rule, the Bayesian estimator yields the most probable value given the observation $\mathbf{y}$ [15]:

$$\begin{aligned} \widehat{\mathbf{h}} &= \arg\max_{\mathbf{h} \in \mathbb{C}^{LM \times 1}} p(\mathbf{h}|\mathbf{y}) \\ &= \arg\min_{\mathbf{h} \in \mathbb{C}^{LM \times 1}} l(\mathbf{h}) \\ &= (\sigma_n^2 \mathbf{I}_{LM} + \mathbf{R}\tilde{\mathbf{S}}^H \tilde{\mathbf{S}})^{-1} \mathbf{R}\tilde{\mathbf{S}}^H \mathbf{y}. \end{aligned} \quad (13)$$

Interestingly, the Bayesian estimate as shown in (13) coincides with the minimum mean square error (MMSE) estimate, which has the form

$$\widehat{\mathbf{h}}^{\text{MMSE}} = \mathbf{R}\tilde{\mathbf{S}}^H (\tilde{\mathbf{S}}\mathbf{R}\tilde{\mathbf{S}}^H + \sigma_n^2 \mathbf{I}_{\tau M})^{-1} \mathbf{y}. \quad (14)$$

(13) and (14) are equivalent thanks to the matrix inversion identity $(\mathbf{I} + \mathbf{AB})^{-1}\mathbf{A} = \mathbf{A}(\mathbf{I} + \mathbf{BA})^{-1}$.

*C. Channel Estimation with Full Pilot Reuse*

Previously we have given expressions whereby interfering channels are estimated simultaneously with the desired channel. This could be of use in designing zero-forcing type receivers. Even though it is clear that Zero-Forcing (ZF) type (or other sophisticated) receivers would give better performance at finite $M$ (see [3] for an analysis of this problem), in this paper, however, we focus on simple matched filters, since such filters are made more relevant by the users of massive MIMO. Matched filters require the knowledge of the desired channel only, so that interference channels can be considered as nuisance parameters. For this case, the single user channel estimation shown below can be used. For ease of exposition, the worst case situation with an unique pilot sequence reused in all $L$ cells is considered:

$$\mathbf{s} = \begin{bmatrix} s_1 & s_2 & \cdots & s_\tau \end{bmatrix}^T. \quad (15)$$

Similar to (7), we define a training matrix $\bar{\mathbf{S}} \triangleq \mathbf{s} \otimes \mathbf{I}_M$. Note that $\bar{\mathbf{S}}^H \bar{\mathbf{S}} = \tau \mathbf{I}_M$. Then the vectorized received training signal at the target base station can be expressed as

$$\mathbf{y} = \bar{\mathbf{S}} \sum_{l=1}^{L} \mathbf{h}_l + \mathbf{n}. \quad (16)$$

Since the Bayesian estimator and the MMSE estimator are identical, we omit the derivation and simply give the expression of this estimator for the desired channel $\mathbf{h}_1$ only:

$$\widehat{\mathbf{h}}_1 = \mathbf{R}_1 \bar{\mathbf{S}}^H \left( \bar{\mathbf{S}} \left( \sum_{l=1}^{L} \mathbf{R}_l \right) \bar{\mathbf{S}}^H + \sigma_n^2 \mathbf{I}_{\tau M} \right)^{-1} \mathbf{y} \quad (17)$$

$$= \mathbf{R}_1 \left( \sigma_n^2 \mathbf{I}_M + \tau \sum_{l=1}^{L} \mathbf{R}_l \right)^{-1} \bar{\mathbf{S}}^H \mathbf{y}. \quad (18)$$

Note that the MMSE channel estimation in the presence of identical pilots is also undertaken in other works such as [3].

In the section below, we examine the degradation caused by the pilot contamination on the estimation performance. In particular, we point out the role played by the use of covariance matrices in dramatically reducing the pilot contamination effects under certain conditions on the rank structure.

We are interested in the mean squared error (MSE) of the proposed estimators, which can be defined as: $\mathcal{M} \triangleq \mathbb{E}\{\|\widehat{\mathbf{h}} - \mathbf{h}\|_F^2\}$, or for the single user channel estimate $\mathcal{M}_1 \triangleq \mathbb{E}\{\|\widehat{\mathbf{h}}_1 - \mathbf{h}_1\|_F^2\}$.

The estimation MSE of (13) is

$$\mathcal{M} = \mathrm{tr}\left\{ \mathbf{R} \left( \mathbf{I}_{LM} + \frac{\tilde{\mathbf{S}}^H \tilde{\mathbf{S}}}{\sigma_n^2} \mathbf{R} \right)^{-1} \right\}. \quad (19)$$

Specifically, when identical pilots are used in all cells, the MSEs are

$$\mathcal{M} = \mathrm{tr}\left\{ \mathbf{R} \left( \mathbf{I}_{LM} + \frac{\tau \mathbf{J}_{LL} \otimes \mathbf{I}_M}{\sigma_n^2} \mathbf{R} \right)^{-1} \right\}, \quad (20)$$

$$\mathcal{M}_1 = \mathrm{tr}\left\{ \mathbf{R}_1 - \mathbf{R}_1^2 \left( \frac{\sigma_n^2}{\tau} \mathbf{I}_M + \sum_{l=1}^{L} \mathbf{R}_l \right)^{-1} \right\}, \quad (21)$$

where $\mathbf{J}_{LL}$ is a $L \times L$ unit matrix consisting of all 1s. The derivations to obtain $\mathcal{M}$ and $\mathcal{M}_1$ use standard methods and the details are omitted here due to lack of space. However, similar methods can be found in [16]. Of course, it is clear from (20) and (21) that the MSE is not dependent on the specific design of the pilot sequence, but on the power of it.

We can readily obtain the channel estimate of (18) in an interference-free scenario, by setting interference terms to zero:

$$\widehat{\mathbf{h}}_1^{no\ int} = \mathbf{R}_1 \left( \sigma_n^2 \mathbf{I}_M + \tau \mathbf{R}_1 \right)^{-1} \bar{\mathbf{S}}^H (\bar{\mathbf{S}} \mathbf{h}_1 + \mathbf{n}), \quad (22)$$

where the superscript *no int* refers to the "no interference case", and the corresponding MSE:

$$\mathcal{M}_1^{no\ int} = \mathrm{tr}\left\{ \mathbf{R}_1 \left( \mathbf{I}_M + \frac{\tau}{\sigma_n^2} \mathbf{R}_1 \right)^{-1} \right\}. \quad (23)$$

### D. Large Scale Analysis

We seek to analyze the performance for the above estimators in the regime of large antenna number $M$. For tractability, our analysis is based on the assumption of an uniform linear array (ULA) with supercritical antenna spacing (i.e., less than or equal to half wavelength).

Hence we have the following multipath model[3]

$$\mathbf{h}_i = \frac{1}{\sqrt{P}} \sum_{p=1}^{P} \mathbf{a}(\theta_{ip}) \alpha_{ip}, \quad (24)$$

where $P$ is the arbitrary number of i.i.d. paths, $\alpha_{ip} \sim \mathcal{CN}(0, \delta_i^2)$ is independent over channel index $i$ and path index $p$, where $\delta_i$ is the $i$-th channel's average attenuation. $\mathbf{a}(\theta)$ is the steering vector, as shown in [17]

$$\mathbf{a}(\theta) \triangleq \begin{bmatrix} 1 \\ e^{-j2\pi \frac{D}{\lambda} \cos(\theta)} \\ \vdots \\ e^{-j2\pi \frac{(M-1)D}{\lambda} \cos(\theta)} \end{bmatrix}, \quad (25)$$

where $D$ is the antenna spacing at the base station and $\lambda$ is the signal wavelength, such that $D \leq \lambda/2$. $\theta_{ip} \in [0, \pi]$ is a random AOA. Note that we can limit angles to $[0, \pi]$ because any $\theta \in [-\pi, 0]$ can be replaced by $-\theta$ giving the same steering vector.

Below, we momentarily assume that the selected users exhibit multipath AOAs that do not overlap with the AOAs for the desired user, i.e., the AOA spread and user locations are such that multipath for the desired user are confined to a region of space where interfering paths are very unlikely to exist. Although the asymptotic analysis below makes use of this condition, it will be shown in Section IV how such a structure can be shaped implicitly by the coordinated pilot assignment. Finally, simulations reveal in Section V the robustness with respect to an overlap between AOA regions of desired and interference channels (for instance in the case of Gaussian AOA distribution).

Our main result is as follows:

**Theorem 1.** *Assume the multipath angle of arrival $\theta$ yielding channel $\mathbf{h}_j, j = 1, \ldots, L$, in (24), is distributed according to an arbitrary density $p_j(\theta)$ with bounded support, i.e., $p_j(\theta) = 0$ for $\theta \notin [\theta_j^{\min}, \theta_j^{\max}]$ for some fixed $\theta_j^{\min} \leq \theta_j^{\max} \in [0, \pi]$. If the $L - 1$ intervals $[\theta_i^{\min}, \theta_i^{\max}]$, $i = 2, \ldots, L$ are strictly non-overlapping with the desired channel's AOA interval*[4] *$[\theta_1^{\min}, \theta_1^{\max}]$, we have*

$$\lim_{M \to \infty} \widehat{\mathbf{h}}_1 = \widehat{\mathbf{h}}_1^{no\ int}. \quad (26)$$

*Proof:* From the channel model (24), we get

$$\mathbf{R_i} = \frac{\delta_i^2}{P} \sum_{p=1}^{P} \mathbb{E}\{\mathbf{a}(\theta_{ip})\mathbf{a}(\theta_{ip})^H\} = \delta_i^2 \mathbb{E}\{\mathbf{a}(\theta_i)\mathbf{a}(\theta_i)^H\},$$

where $\theta_i$ has the PDF $p_i(\theta)$ for all $i = 1, \ldots, L$. The proof of Theorem 1 relies on three intermediate lemmas which exploit the eigenstructures of the covariance matrices. The proofs of the lemmas are given in the appendix. The essential ingredient is to exhibit an asymptotic (at large $M$) orthonormal vector basis for $\mathbf{R}_i$ constructed from steering vectors at regularly sampled spatial frequencies.

---

[3]Note that the Gaussian model (2) can well approximate the multipath model (24) as long as there are enough paths. Since the number of elementary path is typically very large, we have $P \gg 1$ this assumption is valid in practice.

[4]This condition is just one example of practical scenario leading to non-overlapping signal subspaces between the desired and the interference covariances, however, more general multipath scenarios could be used.

**Lemma 1.** *Define $\boldsymbol{\alpha}(x) \triangleq [\ 1\ \ e^{-j\pi x}\ \ \cdots\ \ e^{-j\pi(M-1)x}\ ]^T$ and $\mathcal{A} \triangleq \mathrm{span}\{\boldsymbol{\alpha}(x), x \in [-1, 1]\}$. Given $b_1, b_2 \in [-1, 1]$ and $b_1 < b_2$, define $\mathcal{B} \triangleq \mathrm{span}\{\boldsymbol{\alpha}(x), x \in [b_1, b_2]\}$, then*

- $\dim\{\mathcal{A}\} = M$
- $\dim\{\mathcal{B}\} \sim (b_2 - b_1)M/2$ *when $M$ grows large.*

*Proof:* See Appendix A. $\square$

Lemma 1 characterizes the number of dimensions a linear space has, which is spanned by $\boldsymbol{\alpha}(x)$, in which $x$ plays the role of spatial frequency.

**Lemma 2.** *With a bounded support of AOAs, the rank of channel covariance matrix $\mathbf{R}_i$ satisfies:*

$$\frac{\mathrm{rank}(\mathbf{R}_i)}{M} \leqslant d_i, \text{ as } M \to \infty,$$

*where $d_i$ is defined as*

$$d_i \triangleq \left(\cos(\theta_i^{min}) - \cos(\theta_i^{max})\right)\frac{D}{\lambda}.$$

*Proof:* See Appendix B. $\square$

Lemma 2 indicates that for large $M$, there exists a null space $\mathrm{null}(\mathbf{R}_i)$ of dimension $(1 - d_i)M$. Interestingly, related eigenstructure properties of the covariance matrices were independently derived in [18] for the purpose of reducing the overhead of downlink channel estimation and CSI feedback in massive MIMO for FDD systems.

**Lemma 3.** *The null space $\mathrm{null}(\mathbf{R}_i)$ includes a certain set of unit-norm vectors:*

$$\mathrm{null}(\mathbf{R}_i) \supset \mathrm{span}\left\{\frac{\mathbf{a}(\Phi)}{\sqrt{M}}, \forall \Phi \notin [\theta_i^{min}, \theta_i^{max}]\right\}, \text{ as } M \to \infty.$$

*Proof:* See Appendix C. $\square$

This lemma indicates that multipath components with AOA outside the AOA region for a given user will tend to fall in the null space of its covariance matrix in the large-number-of-antennas case.

We now return to the proof of Theorem 1. $\mathbf{R}_i$ can be decomposed into

$$\mathbf{R}_i = \mathbf{U}_i \boldsymbol{\Sigma}_i \mathbf{U}_i^H, \quad (27)$$

where $\mathbf{U}_i$ is the signal eigenvector matrix of size $M \times m_i$, in which $m_i \leq d_i M$. $\boldsymbol{\Sigma}_\mathbf{i}$ is an eigenvalue matrix of size $m_i \times m_i$. Due to Lemma 3 and the fact that densities $p_i(\theta)$ and $p_1(\theta)$ have non-overlapping supports, we have

$$\mathbf{U}_i^H \mathbf{U}_1 = 0, \forall i \neq 1, \text{ as } M \to \infty. \quad (28)$$

Combining the channel estimate (18) and the channel model (16), we obtain

$$\widehat{\mathbf{h}}_1 = \mathbf{R}_1 \left(\sigma_n^2 \mathbf{I}_M + \tau \sum_{l=1}^{L} \mathbf{R}_l\right)^{-1} \bar{\mathbf{S}}^H \left(\bar{\mathbf{S}} \sum_{i=1}^{L} \mathbf{h}_i + \mathbf{n}\right).$$

According to (28), matrices $\mathbf{R}_1$ and $\sum_{l=2}^{L} \mathbf{R}_l$ span orthogonal subspaces in the large $M$ limit. Therefore we place ourselves in the asymptotic regime for $M$, when $\tau \sum_{l=2}^{L} \mathbf{R}_l$ can be eigen-decomposed according to

$$\tau \sum_{l=2}^{L} \mathbf{R}_l = \mathbf{W}\boldsymbol{\Sigma}\mathbf{W}^H, \quad (29)$$

where $\mathbf{W}$ is the eigenvector matrix such that $\mathbf{W}^H\mathbf{W} = \mathbf{I}$ and $\mathrm{span}\{\mathbf{W}\}$ is included in the orthogonal complement of $\mathrm{span}\{\mathbf{U}_1\}$. Now denote $\mathbf{V}$ the unitary matrix corresponding to the orthogonal complement of both $\mathrm{span}\{\mathbf{W}\}$ and $\mathrm{span}\{\mathbf{U}_1\}$, so that the $M \times M$ identity matrix can now be decomposed into:

$$\mathbf{I}_M = \mathbf{U}_1\mathbf{U}_1^H + \mathbf{W}\mathbf{W}^H + \mathbf{V}\mathbf{V}^H. \quad (30)$$

Thus, for large $M$,

$$\widehat{\mathbf{h}}_1 \sim \mathbf{U}_1\boldsymbol{\Sigma}_1\mathbf{U}_1^H \left(\sigma_n^2 \mathbf{U}_1\mathbf{U}_1^H + \sigma_n^2 \mathbf{V}\mathbf{V}^H + \sigma_n^2 \mathbf{W}\mathbf{W}^H\right.$$
$$\left. + \tau\mathbf{U}_1\boldsymbol{\Sigma}_1\mathbf{U}_1^H + \mathbf{W}\boldsymbol{\Sigma}\mathbf{W}^H\right)^{-1} \left(\tau \sum_{i=1}^{L} \mathbf{h}_i + \bar{\mathbf{S}}^H\mathbf{n}\right).$$

Due to asymptotic orthogonality between $\mathbf{U}_1$, $\mathbf{W}$ and $\mathbf{V}$,

$$\widehat{\mathbf{h}}_1 \sim \mathbf{U}_1\boldsymbol{\Sigma}_1(\sigma^2\mathbf{I}_{m_1} + \tau\boldsymbol{\Sigma}_1)^{-1}\mathbf{U}_1^H(\tau\sum_{i=1}^{L}\mathbf{h}_i + \bar{\mathbf{S}}^H\mathbf{n})$$

$$\sim \mathbf{U}_1\boldsymbol{\Sigma}_1(\sigma^2\mathbf{I}_{m_1} + \tau\boldsymbol{\Sigma}_1)^{-1}\tau(\mathbf{U}_1^H\mathbf{h}_1 + \sum_{i=2}^{L}\mathbf{U}_1^H\mathbf{h}_i + \frac{\bar{\mathbf{S}}^H\mathbf{n}_1}{\tau}).$$

However, since $\mathbf{h}_i \subset \mathrm{span}\left\{\mathbf{a}(\theta), \forall \theta \in [\theta_i^{\min}, \theta_i^{\max}]\right\}$, we have from Lemma 3 that $\frac{\|\mathbf{U}_1^H\mathbf{h}_i\|}{\|\mathbf{U}_1^H\mathbf{h}_1\|} \to 0$, for $i \neq 1$ when $M \to \infty$. Therefore

$$\lim_{M \to \infty} \widehat{\mathbf{h}}_1 = \tau\mathbf{U}_1\boldsymbol{\Sigma}_1\left(\sigma_n^2\mathbf{I}_{m_1} + \tau\boldsymbol{\Sigma}_1\right)^{-1}\left(\mathbf{U}_1^H\mathbf{h}_1 + \frac{\bar{\mathbf{S}}^H\mathbf{n}}{\tau}\right),$$

which is identical to $\widehat{\mathbf{h}}_1^{\text{no int}}$ if we apply the EVD decomposition (27) for $\mathbf{R}_1$ in (22). This proves Theorem 1. $\square$

We also believe that, although antenna calibration is needed as a technical assumption in the theorem, orthogonality of covariances signal subspaces will occur in non-tightly calibrated settings provided the AOA regions do not overlap.

## IV. COORDINATED PILOT ASSIGNMENT

We have seen from above that the performance of the covariance-aided channel estimation is particularly sensitive to the degree with which the signal subspaces of covariance matrices for the desired and the interference channels overlap with each other. In the ideal case where the desired and the interference covariances span distinct subspaces, we have demonstrated that the pilot contamination effect tends to vanish in the large-antenna-array case. In this section, we make use of this property by designing a suitable coordination protocol for assigning pilot sequences to users in the $L$ cells. The role of the coordination is to optimize the use of covariance matrices in an effort to try and satisfy the non-overlapping AOA constraint of Theorem 1. We assume that in all $L$ cells, the considered pilot sequence will be assigned to one (out of $K$) user in each of the $L$ cells. Let

$\mathcal{G} \triangleq \{1, \ldots, K\}$, then $\mathcal{K}_l \in \mathcal{G}$ denotes the index of the user in the $l$-th cell who is assigned the pilot sequence $\mathbf{s}$. The set of selected users is denoted by $\mathcal{U}$ in what follows.

We use the estimation MSE (21) as a performance metric to be minimized in order to find the best user set. (20) is an alternative if we take the estimates of interfering channels into consideration. For a given user set $\mathcal{U}$, we define a network utility function

$$\mathrm{F}(\mathcal{U}) \triangleq \sum_{j=1}^{|\mathcal{U}|} \frac{\mathcal{M}_j(\mathcal{U})}{\mathrm{tr}\{\mathbf{R}_{jj}(\mathcal{U})\}}, \quad (31)$$

where $|\mathcal{U}|$ is the cardinal number of the set $\mathcal{U}$. $\mathcal{M}_j(\mathcal{U})$ is the estimation MSE for the desired channel at the $j$-th base station, with a notation readily extended from $\mathcal{M}_1$ in (21), where this time cell $j$ is the target cell when computing $\mathcal{M}_j$. $\mathbf{R}_{jj}(\mathcal{U})$ is the covariance matrix of the desired channel at the $j$-th cell.

The principle of the coordinated pilot assignment consists in exploiting covariance information at all cells (a total of $KL^2$ covariance matrices) in order to minimize the sum MSE metric. Hence, $L$ users are assigned an identical pilot sequence when the corresponding $L^2$ covariance matrices exhibit the most orthogonal signal subspaces. Note that the MSE-based criterion (31) implicitly exploits the property of subspace orthogonality, e.g., at high SNRs, the proposed MSE-based criterion will be minimized by choices of users with covariance matrices showing maximum signal subspace orthogonality, thereby implicitly satisfying the conditions behind Theorem 1. In view of minimizing the search complexity, a classical greedy approach is proposed:

1) Initialize $\mathcal{U} = \emptyset$
2) For $l = 1, \ldots, L$ do:
   $\mathcal{K}_l = \arg\min_{k \in \mathcal{G}} \mathrm{F}(\mathcal{U} \cup \{k\})$
   $\mathcal{U} \leftarrow \mathcal{U} \cup \{\mathcal{K}_l\}$
End

The coordination can be interpreted as follows: To minimize the estimation error, a base station tends to assign a given pilot to the user whose spatial feature has most differences with the interfering users assigned the same pilot. Clearly, the performance will improve with the number of users, as it becomes more likely to find users with discriminable second-order statistics.

## V. NUMERICAL RESULTS

In order to preserve fairness between users and avoid having high-SNR users being systematically assigned the considered pilot, we consider a symmetric multicell network where the users are all distributed on the cell edge and have the same distance with their base stations. In practice, users with greater average SNR levels (but equal across cells) can be assigned together on a separate pilot pool. We adopt the model of a cluster of synchronized and hexagonally shaped cells. Some basic simulation parameters are given in Table I. We keep these parameters in the following simulation, unless otherwise stated.

TABLE I
BASIC SIMULATION PARAMETERS

| Cell radius | 1 km |
|---|---|
| Cell edge SNR | 20 dB |
| Number of user per-cell | 10 |
| Distance from a user to its BS | 800 m |
| Path loss exponent | 3 |
| Carrier frequency | 2 GHz |
| Antenna spacing | $\lambda/2$ |
| Number of paths | 50 |
| Pilot length | 10 |

The channel vector between the $u$-th user in the $l$-th cell and the target base station is

$$\mathbf{h}_{lu} = \frac{1}{\sqrt{P}} \sum_{p=1}^{P} \mathbf{a}(\theta_{lup}) \alpha_{lup}, \quad (32)$$

where $\theta_{lup}$ and $\alpha_{lup}$ are the AOA and the attenuation of the $p$-th path between the $u$-th user in the $l$-th cell and the target base station respectively. Note that the variance of $\alpha_{lup}, \forall p$ is $\delta_{lu}^2$, which includes the distance-based path loss $\beta_{lu}$ between the user and the target base station (which can be anyone of the $L$ cells):

$$\beta_{lu} = \frac{\alpha}{d_{lu}^\gamma}, \quad (33)$$

where $\alpha$ is a constant dependent on the prescribed average SNR at cell edge. $d_{lu}$ is the geographical distance. $\gamma$ is the path-loss exponent.

Two types of AOA distributions are considered here, a non-bounded one (Gaussian) and a bounded one (uniform):

*1) Gaussian distribution:* For the channel coefficients $\mathbf{h}_{lu}$, the AOAs of all $P$ paths are i.i.d. Gaussian random variables with mean $\bar{\theta}_{lu}$ and standard deviation $\sigma$. Here we suppose all the desired channels and interference channels have the same standard deviation of AOA. Note that Gaussian AOA distributions cannot fulfill the conditions of non-overlapping AOA support domains in Theorem 1, nevertheless the use of the proposed method in this context also gives substantial gains as $\sigma^2$ decreases.

*2) Uniform distribution:* For the channel $\mathbf{h}_{lu}$, the AOAs are uniformly distributed over $[\bar{\theta}_{lu} - \theta_\Delta, \bar{\theta}_{lu} + \theta_\Delta]$, where $\bar{\theta}_{lu}$ is the mean AOA.

Two performance metrics are used to evaluate the proposed channel estimation scheme. The first one is a normalized channel estimation error

$$\mathrm{err} \triangleq 10\log_{10}\left(\frac{\sum_{j=1}^{L} \left\|\widehat{\mathbf{h}}_{jj} - \mathbf{h}_{jj}\right\|_F^2}{\sum_{j=1}^{L} \|\mathbf{h}_{jj}\|_F^2}\right), \quad (34)$$

where $\mathbf{h}_{jj}$ and $\widehat{\mathbf{h}}_{jj}$ are the desired channel at the $j$-th base station and its estimate respectively. Note that we only consider the estimation error of the desired channel. The second performance metric is the per-cell rate of the downlink obtained assuming standard MRC beamformer based on the channel estimates. The beamforming weight vector of the $j$-th base station is $\mathbf{w}_j^{\text{MRC}} = \widehat{\mathbf{h}}_{jj}$. We define the per-cell rate as follows:

$$\mathcal{C} \triangleq \frac{\sum_{j=1}^{L} \log_2(1 + \text{SINR}_j)}{L},$$

where $\text{SINR}_j$ is the received signal-to-noise-plus-interference ratio (SINR) by the scheduled user in the $j$-th cell.

Numerical results of the proposed channel estimation scheme are now shown. In the figures, "LS" stands for conventional LS channel estimation. "CB" denotes the Covariance-aided Bayesian estimation (without coordinated pilot assignment), and "CPA" is the proposed Coordinated Pilot Assignment-based Bayesian estimation.

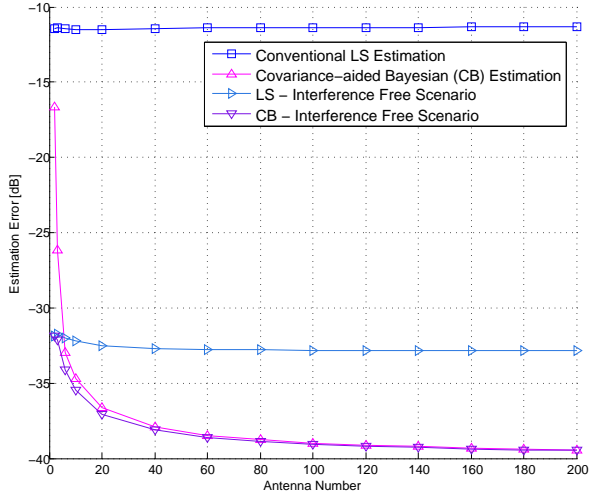

Fig. 1. Estimation MSE vs. BS antenna number, 2-cell network, fixed positions of two users, uniformly distributed AOAs with $\theta_\Delta = 20$ degrees, non-overlapping multipath.

We first validate Theorem 1 in Fig. 1 with a 2-cell network, where the two users' positions are fixed. AOAs of desired channels are uniformly distributed with a mean of 90 degrees, and the angle spreads of all channels are 20 degrees, yielding no overlap between desired and interfering multipaths. The pilot contamination is quickly eliminated with growing number of antennas.

In Fig. 2 and Fig. 3, the estimation MSEs versus the BS antenna number are illustrated. When the AOAs have uniform distributions with $\theta_\Delta = 10$ degrees, as shown in Fig. 2, the performance of CPA estimator improves quickly with $M$ from 2 to 10. In the 2-cell network, CPA has the ability of avoiding the overlap between AOAs for the desired and interference channels. For comparison, Fig. 3 is obtained with Gaussian AOA distribution. We can observe a gap remains between the CPA and the interference-free one, due to the non-boundedness of the Gaussian PDF. Nevertheless, the gains over the classical estimator remain substantial.

We then examine the impact of standard deviation $\sigma$ of Gaussian AOAs on the estimation. Fig. 4 shows that the estimation error is a monotonically increasing function of $\sigma$. In contrast, an angle spread tending toward zero will cause the channel direction to collapse into a deterministic quantity, yielding large gains for covariance-based channel estimation.

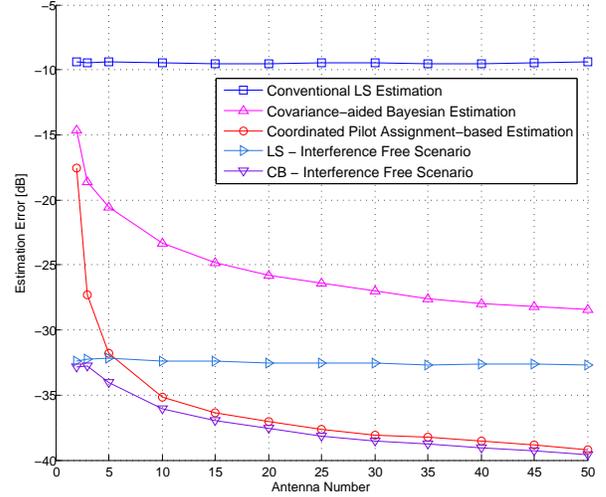

Fig. 2. Estimation MSE vs. antenna number, uniformly distributed AOAs with $\theta_\Delta = 10$ degrees, 2-cell network.

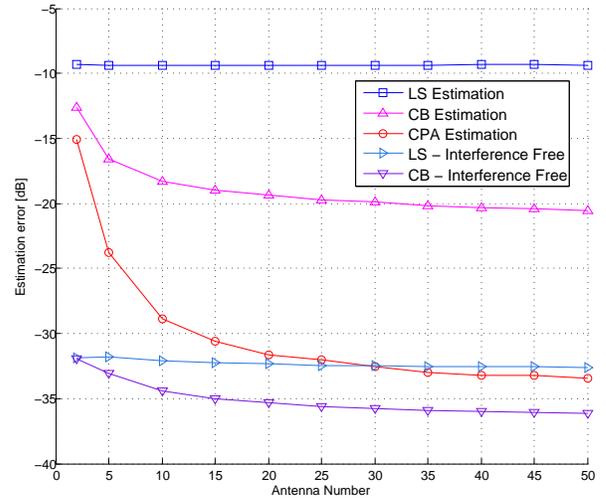

Fig. 3. Estimation MSE vs. antenna number, Gaussian distributed AOAs with $\sigma = 10$ degrees, 2-cell network.

Figs. 5 and 6 depict the downlink per-cell rate achieved by the MRC beamforming strategy and suggest large gains when the Bayesian estimation is used in conjunction with the proposed coordinated pilot assignment strategy and intermediate gains when it is used alone. Obviously the rate performance almost saturates with $M$ in the classical LS case (due to pilot contamination) while it increases quickly with $M$ for the proposed estimators, indicating the full benefits of massive MIMO systems are exploited.

## VI. Discussions

In this paper, we assumed the individual covariance matrices can be estimated separately. This could be done in practice by exploiting resource blocks where the desired user and interference users are known to be assigned at different times. In future networks, one may imagine a specific training design for learning second-order statistics. Since covariance information varies much slower than fast fading, such training may not consume a substantial amount of resources.

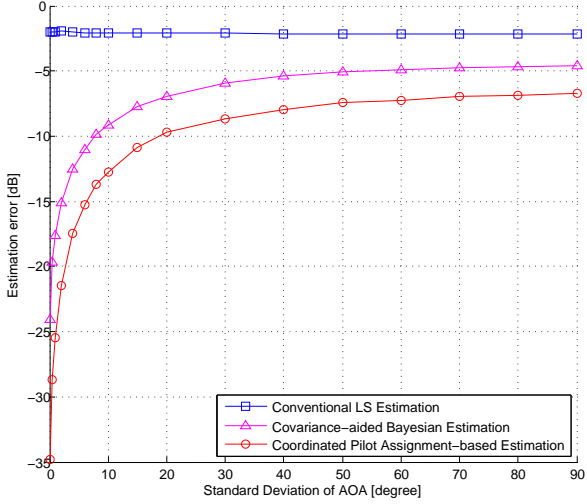

Fig. 4. Estimation MSE vs. standard deviation of Gaussian distributed AOAs with $M = 10$, 7-cell network.

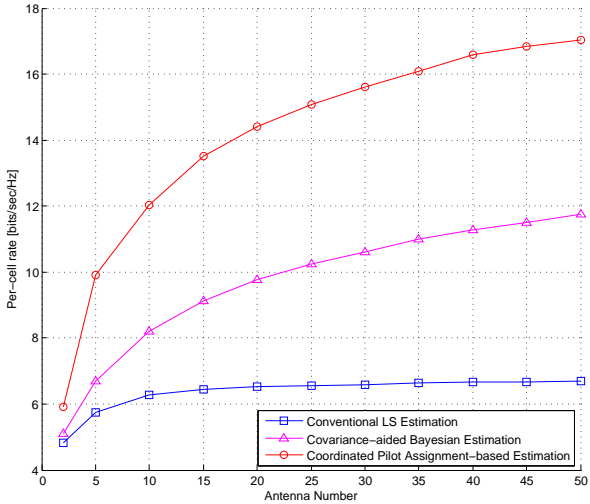

Fig. 5. Per-cell rate vs. antenna number, 2-cell network, Gaussian distributed AOAs with $\sigma = 10$ degrees.

The proposed coordinated estimation method would introduce information exchange between base stations. Although the second-order statistics vary much slower than the instantaneous CSI, base stations still have to update the covariance information every now and then so as to maintain performance. Clearly, the overhead depends on the degree of user mobility.

## VII. CONCLUSIONS

This paper proposes a covariance-aided channel estimation framework in the context of interference-limited multi-cell multiple antenna systems. We develop Bayesian estimators and demonstrate analytically the efficiency of such an approach for large-scale antenna systems, leading to a complete removal of pilot contamination effects in the case covariance matrices satisfy a certain non-overlapping condition on their dominant subspaces. We suggest a coordinated pilot assignment strategy that helps shape covariance matrices toward satisfying the needed condition and show channel estimation performance close to interference-free scenarios.

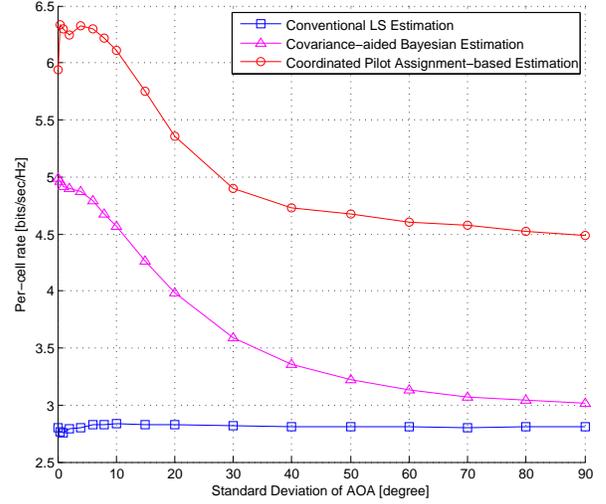

Fig. 6. Per-cell rate vs. standard deviation of AOA (Gaussian distribution) with $M = 10$, 7-cell network.

## APPENDIX

### A. Proof of Lemma 1:

Define the series
$$x_i \triangleq -1 + \frac{2(i-1)}{M}, i = 1, \ldots, M,$$
and
$$\boldsymbol{\mu}_i \triangleq \frac{\boldsymbol{\alpha}(x_i)}{\sqrt{M}}.$$

Then we have $\boldsymbol{\mu}_i \subset \mathcal{A}, \forall i = 1, \ldots, M$ and
$$\boldsymbol{\mu}_k^H \boldsymbol{\mu}_i = \frac{1 - e^{-j2\pi(i-k)}}{M(1 - e^{-\frac{j2\pi(i-k)}{M}})} = 0, k \neq i.$$

Thus $\{\boldsymbol{\mu}_i \mid i = 1, \ldots, M\}$ forms an orthogonal basis of $\mathcal{A}$, and therefore
$$\dim\{\mathcal{A}\} = M.$$

Define
$$\tilde{B} \triangleq \left\{ \boldsymbol{\mu}_i \,\middle|\, i \in \mathbb{Z} \cap \left[\lfloor \frac{M(b_1+1)}{2} + 1 \rfloor + 1, \lceil \frac{M(b_2+1)}{2} + 1 \rceil - 1\right] \right\},$$

where $\lceil x \rceil$ and $\lfloor x \rfloor$ are rounded-above and rounded-below operators respectively. Then $\tilde{B}$ is part of an orthogonal basis of the space $\mathcal{B}$, which indicates $\dim\{\mathcal{B}\} \geq |\tilde{B}|$. By counting vectors in $\tilde{B}$, we have that

$$\begin{aligned}\dim\{\mathcal{B}\} &\geq \lceil \frac{M(b_2+1)}{2} + 1 \rceil - \lfloor \frac{M(b_1+1)}{2} + 1 \rfloor - 1 \\ &= \lceil \frac{M(b_2+1)}{2} \rceil - \lfloor \frac{M(b_1+1)}{2} \rfloor - 1. \end{aligned} \quad (35)$$

Now we define
$$\tilde{C} \triangleq \left\{ \boldsymbol{\mu}_i \,\middle|\, i \in \mathbb{Z} \text{ and } i \in \left[1, \lfloor \frac{M(b_1+1)}{2} + 1 \rfloor\right] \right. \\ \left. \cup \left[\lceil \frac{M(b_2+1)}{2} + 1 \rceil, M\right]\right\}.$$

Then $\tilde{C}$ is part of an orthogonal basis of $\mathcal{A}$. Furthermore,

$$|\tilde{C}| = \lfloor \frac{M(b_1+1)}{2} + 1 \rfloor + M - \lceil \frac{M(b_2+1)}{2} + 1 \rceil + 1$$
$$= M - \lceil \frac{M(b_2+1)}{2} \rceil + \lfloor \frac{M(b_1+1)}{2} \rfloor + 1.$$

Consider the equivalent form of $\mathcal{B}$

$$\mathcal{B} = \left\{ \int_{b_1}^{b_2} f(x)\boldsymbol{\alpha}(x)dx \,\bigg|\, \forall |f(x)| < \infty,\, x \in [b_1, b_2] \right\}.$$

Take any vector $\boldsymbol{\mu}_i \in \tilde{C}$, we have

$$\boldsymbol{\mu}_i^H \int_{b_1}^{b_2} f(x)\boldsymbol{\alpha}(x)dx = \frac{1}{\sqrt{M}} \int_{b_1}^{b_2} f(x)\boldsymbol{\alpha}(x_i)^H \boldsymbol{\alpha}(x) dx$$
$$= \frac{1}{\sqrt{M}} \int_{b_1}^{b_2} f(x) \frac{1 - e^{-j\pi M(x-x_i)}}{1 - e^{-j\pi(x-x_i)}} dx.$$

Since $\boldsymbol{\mu}_i \in \tilde{C}$, we can observe $x_i \notin [b_1, b_2]$, thus

$$\lim_{M \to \infty} \boldsymbol{\mu}_i^H \int_{b_1}^{b_2} f(x)\boldsymbol{\alpha}(x) dx = 0.$$

Therefore $\tilde{C} \subset \mathcal{B}^\perp$ when $M \to \infty$. Hence we have

$$\dim\{\mathcal{B}^\perp\} = M - \dim\{\mathcal{B}\} \geqslant |\tilde{C}|$$
$$\Rightarrow \dim\{\mathcal{B}\} \leqslant \lceil \frac{M(b_2+1)}{2} \rceil - \lfloor \frac{M(b_1+1)}{2} \rfloor - 1. \quad (36)$$

Combining (35) and (36), we can easily obtain

$$\dim\{\mathcal{B}\} \sim \lceil \frac{M(b_2+1)}{2} \rceil - \lfloor \frac{M(b_1+1)}{2} \rfloor - 1$$
$$\sim \frac{M(b_2-b_1)}{2} + o(M),$$

and Lemma 1 is proved. □

### B. Proof of Lemma 2:

We define

$$\mathbf{b}(x) \triangleq \mathbf{a}\left(\cos^{-1}(x\frac{\lambda}{D})\right), x \in [-\frac{D}{\lambda}, \frac{D}{\lambda}]. \quad (37)$$

It is clear from Lemma 1 that $\mathbf{b}(x) = \boldsymbol{\alpha}(2x)$. Hence, for any interval $[x^{\min}, x^{\max}]$ in $[-\frac{1}{2}, \frac{1}{2}]$,

$$\dim\left\{\text{span}\left\{\mathbf{b}(x), \forall x \in [x^{\min}, x^{\max}]\right\}\right\}$$
$$\sim (x^{\max} - x^{\min})M \text{ when } M \text{ is large.} \quad (38)$$

Additionally, for $i = 1, \ldots, L$, we have

$$\text{span}\{\mathbf{R}_i\} = \text{span}\left\{\int_0^\pi \mathbf{a}(\theta)\mathbf{a}(\theta)^H p_i(\theta) d\theta\right\},$$

Thus, due to the bounded support of $p_i(\theta)$, we can obtain

$$\text{span}\{\mathbf{R}_i\} = \text{span}\left\{\int_{\theta_i^{\min}}^{\theta_i^{\max}} \mathbf{a}(\theta)\mathbf{a}(\theta)^H p_i(\theta) d\theta\right\}$$
$$= \text{span}\left\{\int_{\theta_i^{\min}}^{\theta_i^{\max}} \mathbf{b}(\frac{D}{\lambda}\cos(\theta))\mathbf{b}^H(\frac{D}{\lambda}\cos(\theta)) p_i(\theta) d\theta\right\}.$$

Then, by interpreting the integral as a (continuous) sum, we have

$$\text{span}\{\mathbf{R}_i\} \subset \text{span}\left\{\mathbf{b}(x), \forall x \in [\frac{D}{\lambda}\cos(\theta_i^{\max}), \frac{D}{\lambda}\cos(\theta_i^{\min})]\right\}.$$

From (38), we obtain

$$\text{rank}(\mathbf{R}_i) \leq \left(\cos(\theta_i^{\min}) - \cos(\theta_i^{\max})\right)\frac{D}{\lambda}M,$$

for large $M$, and Lemma 2 is proved. □

### C. Proof of Lemma 3:

Take an angle $\Phi \notin [\theta_i^{\min}, \theta_i^{\max}]$ and define

$$\mathbf{u} \triangleq \frac{\mathbf{a}(\Phi)}{\sqrt{M}}.$$

Then we have

$$\mathbf{u}^H \mathbf{R}_i \mathbf{u} = \frac{1}{M}\mathbf{a}(\Phi)^H \mathbf{R}_i \mathbf{a}(\Phi)$$
$$= \frac{1}{M}\mathbf{a}^H(\Phi)\mathbb{E}\left\{\mathbf{a}(\theta)\mathbf{a}^H(\theta)\right\}\mathbf{a}(\Phi)$$
$$= \frac{1}{M}\mathbb{E}\left\{\left|\mathbf{a}^H(\Phi)\mathbf{a}(\theta)\right|^2\right\}$$
$$= \frac{1}{M}\mathbb{E}\left\{\left|\sum_{m=0}^{M-1} e^{2\pi j(m-1)\frac{D}{\lambda}(\cos(\Phi)-\cos(\theta))}\right|^2\right\}$$
$$= \int_{\theta_i^{\min}}^{\theta_i^{\max}} \left|\frac{1}{M}\sum_{m=0}^{M-1} e^{2\pi j(m-1)\frac{D}{\lambda}(\cos(\Phi)-\cos(\theta))}\right|^2 p_i(\theta) d\theta.$$

According to the well-known result on the sum of geometric series, we can easily obtain

$$\lim_{M \to \infty} \left|\frac{1}{M}\sum_{m=0}^{M-1} e^{2\pi j(m-1)\frac{D}{\lambda}(\cos(\Phi)-\cos(\theta))}\right|^2 = 0,$$

since $\Phi \neq \theta, \forall \theta \in [\theta_i^{\min}, \theta_i^{\max}]$. Thus

$$\lim_{M \to \infty} \mathbf{u}^H \mathbf{R}_i \mathbf{u} = 0,$$

which proves Lemma 3. □

### ACKNOWLEDGMENT

Discussions with Dirk Slock and Laura Cottatellucci are gratefully acknowledged.